\documentclass[8pt,twocolumn]{article}

\usepackage[matrix,frame,arrow]{xy}
\usepackage{amsmath}

\newcommand{\ket}[1]{\left\vert{#1}\right\rangle}

\newcommand{\qw}[1][-1]{\ar @{-} [0,#1]}

\newcommand{\cw}[1][-1]{\ar @{=} [0,#1]}

\newcommand{\nwx}[1][-1]{\ar @{.} [#1,0]} 

\newcommand{\gate}[1]{*{\xy *+<.6em>{#1};p\save+LU;+RU **\dir{-}\restore\save+RU;+RD **\dir{-}\restore\save+RD;+LD **\dir{-}\restore\POS+LD;+LU **\dir{-}\endxy} \qw}
\newcommand{\meter}{\gate{\xy *!<0em,1.1em>h\cir<1.1em>{ur_dr},!U-<0em,.4em>;p+<.5em,.9em> **h\dir{-} \POS <-.6em,.4em> *{},<.6em,-.4em> *{} \endxy}}

\newcommand{\multigate}[2]{*+<1em,.9em>{\hphantom{#2}} \qw \POS[0,0].[#1,0];p !C *{#2},p \save+LU;+RU **\dir{-}\restore\save+RU;+RD **\dir{-}\restore\save+RD;+LD **\dir{-}\restore\save+LD;+LU **\dir{-}\restore}
\newcommand{\ghost}[1]{*+<1em,.9em>{\hphantom{#1}} \qw}

\newcommand{\lstick}[1]{*!R!<.5em,0em>=<0em>{#1}}

\newcommand{\Qcircuit}[1][0em]{\xymatrix @*[o] @*=<#1>}

\usepackage{graphicx,url}
\usepackage[english]{babel}
\usepackage[latin1]{inputenc}
\usepackage{amssymb}
\usepackage{amsmath}
\usepackage{algorithmic}
\usepackage{float}
\floatstyle{ruled}
\newfloat{program}{thp}{lop}
\floatname{program}{Pseudocode}

\title{Examples of the Generalized Quantum Permanent Compromise Attack to the Blum-Micali Construction}

\author{Elloá B. Guedes, Francisco M. de Assis, Bernardo Lula Jr.\\ IQuanta -- Institute for Studies in Quantum Computation and Quantum Information\\
   Federal University of Campina Grande\\
  Rua Aprígio Veloso, 882 -- Campina Grande -- Paraíba -- Brazil\\
    \texttt{elloaguedes@gmail.com, fmarcos@dee.ufcg.edu.br, lula@dsc.ufcg.edu.br}
}

\begin{document}
\selectlanguage{english}

\maketitle

\section*{File Description}
This file contains examples of the generalized quantum permanent compromise attack to the Blum-Micali construction. The examples presented here illustrate the attack described in the paper published by Guedes et al. in WECIQ $2010$ [3].

To characterize the Blum-Blum-Shub generator, the following references were used: [1, 5, 8, 10] . In the case of the Kaliski generator, the references were: [2, 6, 8, 10]. The reader should consulte them to see more details about these generators.


\section{Blum-Blum-Shub Generator}
Let $M$ be the product of two large primes $p$ and $q$ where $p \equiv q \equiv 3 \bmod 4$, i.e., $M$ is a Blum prime. Define $QR_M$ as the quadratic residues modulo $M$, i.e., $QR_M = (\mathbb{Z}_M^{*})^2$.

Let $f: \mathbb{Z}_M \rightarrow \mathbb{Z}_M$ be the Rabin function, with the following definition

\begin{equation}
f(x) = x^2 \bmod M
\end{equation}

The Blum-Blum-Shub generator (BBS) takes $x_0 \in_R \mathbb{Z}_{M}^{*}$ and iterates the Rabin function in the following way:

\begin{eqnarray}
x_{i} &=& x_{i-1}^2 \bmod M\\
b_i &=& \gamma_j(x_i)
\end{eqnarray} where $\gamma_j$ denotes the hard-core predicate for the one-way permutation. This hard-core predicate returns the $j$-th bit from the given parameter, where $j$ is previously fixed and $1 < j < n$. The value of $M$ and $j$ are publicly know and the security of the BBS generator relies on the hypothesis of the hardness of factoring [5, 8, 10].

Suppose that a cryptosystem uses the BBS to produce pseudorandom quantities. This generator was initialized with the parameters $(M = 3 \cdot 7 = 21, j = 5)$ that are publicly known\footnote{Considering $j=5$ represents that the least significant bit will be returned by the hard-core predicate.}.

Suppose that an adversary of this cryptosystem wants to attack the BBS generator. In this scenario, suppose that the adversary ($i$) discovered that the following sequence of bits $\textbf{b} = 10$ was outputted by the generator; and, ($ii$) possess a quantum computer able to execute the generalized quantum permanent compromise attack to the Blum-Micali construction.

In the next sections, the activities to perform the attack successfully will be described. 

\subsection{Attack Setup}
The attack setup comprehend all the steps necessary to prepare the quantum algorithm to run. Firstly,the adversary needs to prepare the quantum gates that will be used in the attack.

The number of qubits to represent the domain in a quantum computer is $\left\lceil \log \mathcal{D} \right\rceil = 5$. Since $2$ bits where discovered by the adversary, $2$ qubits will compose the second register. In this way, the summarization of necessary qubits is: $5$ qubits to first register, $2$ qubits to the second register, and $1$ qubit as ancillary to the amplitude amplification procedure.

The $\rho$ gate implements the permutation over $QR_M$, that performs the following transformations:

\begin{eqnarray}
\ket{x \in QR_M} &\rightarrow& \ket{x^2 \bmod M}\\
\ket{x \not\in QR_M} &\rightarrow& \ket{x}
\end{eqnarray}

To facilitate the notation, let $lsb(x)$ be the function that, given an integer $x$, returns the least significant bit of $x$.

The $\delta_{b_i}$ gates, where $b_i$ represents the associated bit produced, have the following definition:

\begin{displaymath}
\delta_{b_i}\ket{x}\ket{y} = \left\{ \begin{array}{cc}
                        \ket{x}\ket{\overline{y}} & \textrm{if } lsb(x) = b_i \textrm{ and } x \in QR_M\\
                        \ket{x}\ket{y} & \textrm{otherwise}\\
                      \end{array}
\right.
\end{displaymath} In summary, it can be said that the gate $\delta_{b_i}$ inverts the target qubit, when the value of the control qubit would have produced the associated bit $b_i$ according to the hard-core predicate $lsb$.

The last step of the attack setup is to determine how many Grover's iterations will be necessary. In this case, it is expected just a single solution over $N = \left\lceil \log M \right\rceil = 5$ bits of input, i.e., $32$ numbers. So, the number of iterations $k$ is given by:

\begin{equation}
k = \left\lfloor \frac{\pi}{4}\sqrt{\frac{32}{1}}\right\rceil = 4
\end{equation}

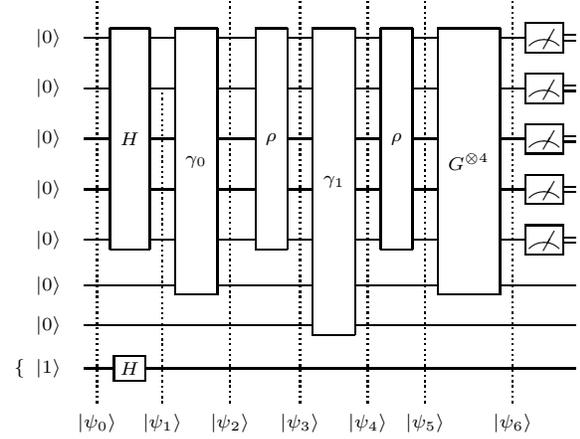
\begin{figure}[ht!]
\centering
\leavevmode
\scriptsize{
\Qcircuit @C=0.6em @R=1.0em {
&&&&&&&&&&&&&&&&&&&&&&&&&&&&&&&&&&&&&&\\
& & & & \lstick{\ket{0}}   & & \qw & \multigate{4}{H} & \qw & \multigate{5}{\gamma_0}   & \qw   & \qw  & \multigate{4}{\rho}   & \qw  &\multigate{6}{\gamma_1}  & \qw  & \multigate{4}{\rho} &\qw & \multigate{5}{G^{\otimes4}}    & \qw & \meter & \cw\\
& & & & \lstick{\ket{0}}   & & \qw & \ghost{H}        & \qw & \ghost{\gamma_0}          & \qw   & \qw  & \ghost{\rho}          & \qw  &\ghost{\gamma_1}         & \qw  & \ghost{\rho}        &\qw & \ghost{G^{\otimes4}}           & \qw & \meter & \cw \\
& & & & \lstick{\ket{0}}   & & \qw & \ghost{H}        & \qw & \ghost{\gamma_0}          & \qw   & \qw  & \ghost{\rho}          & \qw  &\ghost{\gamma_1}         & \qw  & \ghost{\rho}        &\qw & \ghost{G^{\otimes4}}           & \qw & \meter & \cw \\
& & & & \lstick{\ket{0}}   & & \qw & \ghost{H}        & \qw & \ghost{\gamma_0}          & \qw   & \qw  & \ghost{\rho}          & \qw  &\ghost{\gamma_1}         & \qw  & \ghost{\rho}        &\qw & \ghost{G^{\otimes4}}           & \qw & \meter & \cw \\
& & & & \lstick{\ket{0}}   & & \qw & \ghost{H}        & \qw & \ghost{\gamma_0}          & \qw   & \qw  & \ghost{\rho}          & \qw  &\ghost{\gamma_1}         & \qw  & \ghost{\rho}        &\qw & \ghost{G^{\otimes4}}           & \qw & \meter & \cw \\
& & & & \lstick{\ket{0}}   & & \qw & \qw              & \qw & \ghost{\gamma_0}          & \qw   & \qw  & \qw                   & \qw  &\ghost{\gamma_1}         & \qw  & \qw                 &\qw & \ghost{G^{\otimes4}}           & \qw & \qw    & \qw \\
& & & & \lstick{\ket{0}}   & & \qw & \qw              & \qw & \qw                        & \qw   & \qw  & \qw                  & \qw  &\ghost{\gamma_1}         & \qw  & \qw                 &\qw & \qw                            & \qw & \qw    & \qw \\
\left\{ \right. & & & & \lstick{\ket{1}}   & & \qw & \gate{H} & \qw &  \qw               & \qw   & \qw  & \qw                  & \qw  & \qw                     & \qw  & \qw                 &\qw & \qw                            & \qw & \qw    & \qw\\
& & & &                    & &\nwx[-9] &              & \nwx[-7]&  & \nwx[-9]   &      &  & \nwx[-9]  &   & \nwx[-9]  &  & \nwx[-9]  & &  \nwx[-9]                  &  &    &  \\
& & & &                    & &\ket{\psi_0} &              & \ket{\psi_1}&  & \ket{\psi_2}   &      &  & \ket{\psi_3}  &   & \ket{\psi_4}  &  & \ket{\psi_5}  & & \ket{\psi_6}           & &        &  &    &  \\
}}
\caption{Quantum circuit that implements the attack against the BBS generator.} \label{circ:total2}
\end{figure}

Arranging the gates as suggested by the algorithm, the resulting circuit is denoted in the Figure \ref{circ:total2}.

\subsection{Attack Example}
Since the requirements for the attack are prepared, the generalized quantum permanent compromise attack is ready to be executed.

The first step is to prepare the four input registers, as shown in $\ket{\psi_0}$ below:

\begin{equation}
\ket{\psi_0} = \ket{00000}\ket{00}\ket{1}
\end{equation}

A superposition of the input is made to represent all the domain of the generator. The last qubit is also put in superposition because it will be used in the amplitude amplification phase:

\begin{equation}
\ket{\psi_1} = \frac{1}{\sqrt{32}}\left( \sum_{i = 0}^{31} \ket{i} \right) \ket{00} \ket{-}
\end{equation}

Emphasizing the domain $QR_M$, the state $\ket{\psi_1}$ can be rewritten as:

\begin{eqnarray}
\ket{\psi_1'} &=&  \frac{1}{\sqrt{32}}\left( \sum_{i = 0}^{31} \ket{i} \right) \ket{00} \ket{-}\\
&=& \frac{1}{\sqrt{32}} \left( \ket{1} + \ket{4} + \ket{7} + \ket{9} + \ket{15} \right. + \nonumber\\
&+& \left. \ket{16} + \ket{18}\right) \ket{00} \ket{-} + \nonumber\\
&+& \frac{1}{\sqrt{32}} \sum_{i=0, i \not\in QR_M}^{31} \ket{i}\ket{00}\ket{-}
\end{eqnarray}

With the first observed bit $b_1 = 1$, the $\delta_1$ gate will be applied, resulting:

\begin{eqnarray}
\ket{\psi_2} &=& \gamma_1 \ket{\psi_1}\\
&=& \frac{1}{\sqrt{32}} \left( \ket{1} + \ket{7} + \ket{9} + \ket{15} \right)\ket{10}\ket{-} + \nonumber\\
&+& \frac{1}{\sqrt{32}}\left( \ket{4} + \ket{16} + \ket{18} \right) \ket{00} \ket{-} + \nonumber\\
&+& \frac{1}{\sqrt{32}} \sum_{i=0, i \not\in QR_M}^{31} \ket{i}\ket{00}\ket{-}
\end{eqnarray}

Up to this point, the algorithm identify $\hat{X}_1 = \left\{ 1,7,9,15\right\}$ as the potential candidates to the representative. It is important to notice that this identification is just in the quantum level.

The Rabin function, implemented by the $\rho$ gate, must be applied to the input:

\begin{eqnarray}
\ket{\psi_3} &=& \rho\ket{\psi_2}\\
&=& \frac{1}{\sqrt{32}} \left( \ket{1} + \ket{7} + \ket{18} + \ket{15} \right)\ket{10}\ket{-} + \nonumber \\
&+& \frac{1}{\sqrt{32}}\left( \ket{4} + \ket{16} + \ket{9} \right) \ket{00} \ket{-} + \nonumber\\
&+& \frac{1}{\sqrt{32}} \sum_{i=0, i \not\in QR_M}^{31} \ket{i}\ket{00}\ket{-}
\end{eqnarray}

The second bit will be used to determine $\hat{X}_2$:

\begin{eqnarray}
\ket{\psi_4} &=& \gamma_0 \ket{\psi_3}\\
&=& \frac{1}{\sqrt{32}}\ket{18}\ket{11}\ket{-} + \nonumber \\
&+& \frac{1}{\sqrt{32}} \left( \ket{1} + \ket{7} + \ket{15} \right)\ket{10}\ket{-}  + \nonumber \\
&+& \frac{1}{\sqrt{32}}\left( \ket{4} + \ket{16}\right) \ket{01}\ket{-} + \ket{9}\ket{00} \ket{-} + \nonumber\\
&+&  \frac{1}{\sqrt{32}} \sum_{i=0, i \not\in QR_M}^{31} \ket{i}\ket{00}\ket{-}
\end{eqnarray}

It is important to notice that $\hat{X}_2 = \left\{ 9 \right\}$ and the solution is already identified in a quantum level. The next step is to simply obtain $x_3$:

\begin{eqnarray}
\ket{\psi_5} &=& \rho \ket{\psi_4}\\
&=& \frac{1}{\sqrt{32}}\ket{9}\ket{11}\ket{-} + \nonumber\\
&+& \frac{1}{\sqrt{32}} \left( \ket{1} + \ket{7} + \ket{15} \right)\ket{10}\ket{-}  + \nonumber \\
&+& \frac{1}{\sqrt{32}}\left( \ket{16} + \ket{4}\right) \ket{01}\ket{-} + \ket{18}\ket{00} \ket{-}  \nonumber \\
&+& + \frac{1}{\sqrt{32}} \sum_{i=0, i \not\in QR_M}^{31} \ket{i}\ket{00}\ket{-}
\end{eqnarray}

The state $\ket{\psi_5}$ can be written as a partition, where \mbox{$z \ne 11$}:

\begin{eqnarray}
\ket{\psi_5'} &=& \frac{1}{\sqrt{32}} \ket{9}\ket{11}\ket{-} + \sum_{i= 0, i \ne 9}^{31} \ket{i}\ket{z}\ket{-}\\
&=& \frac{1}{\sqrt{32}}\ket{\psi_{x_i}} + \sqrt{\frac{31}{32}}\ket{\psi_{\neg x_i}}
\end{eqnarray} It should be noticed that $\ket{\psi_{x_i}} = \ket{9}\ket{11}\ket{-}$ and $\ket{\psi_{\neg x_i}} = \sum_{i= 0, i \ne 9}^{31} \ket{i}\ket{z}\ket{-}$.

Considering the geometric representation of this state, then:

\begin{equation}
\ket{\psi_5'}  = \sin{\theta} \ket{\psi_{x_i}} + \cos(\theta) \ket{\psi_{\neg x_i}}
\end{equation} where $\sin^2\theta = \frac{1}{32}$ and $\theta \in \left(0,\frac{\pi}{2}\right)$, therefore $\theta = 0.17771$ radians.

The next step is to perform $k  = 4$ Grover iterations, resulting:

{\footnotesize
\begin{eqnarray}
\ket{\psi_6} &=& G^{\otimes 4} \ket{\psi_5}\\
&=& \sin[(2\cdot k + 1)\theta] \ket{\psi_{good}} + \nonumber \\
&+& \cos[(2\cdot k + 1)\theta] \ket{\psi_{bad}}\\
&=& \sin[9 \cdot 0.17771] \ket{\psi_{good}} + \nonumber \\
&+& \cos[9 \cdot 0.17771] \ket{\psi_{bad}}\\
&=& \sin(1.599)\ket{\psi_{good}} + \cos(1.599) \ket{\psi_{bad}}
\end{eqnarray}}

A measurement in the second register will return $9$ with probability of $\left| \sin(1.599)\right|^2 \cong 0.9996$. It means that with just two qubits, the representative of the BBS generator was correctly retrieved with high probability.

This concludes an example of the generalized quantum permanent compromise attack against the security of the BBS generator.

\section{Kaliski Generator}
The Kaliski generator is based on the elliptic curve discrete logarithm problem. Let $p$ be a prime, $p \equiv 2 \bmod 3$, and consider a curve $E(\mathbb{F}_p)$ that consists of points \mbox{$(x,y) \in \mathbb{F}_p \times  \mathbb{F}_p$} such that: \begin{equation}
y^2 = x^3 + c
\end{equation} The points of $E(\mathbb{F}_p)$ together with a point at infinity $\mathcal{O}$ form a cyclic additive group of order $p+1$. Let $Q$ be a generator this group and let $\phi$ be a function with the following definition:

\begin{displaymath}
\phi(P) = \left\{ \begin{array}{cc}
                    y & \textrm{if } P = (x,y)\\
                    p & \textrm{if } P = \mathcal{O}
                  \end{array}
\right.
\end{displaymath} The Kaliski generator's one-way permutation and hard-core predicate are given below:

\begin{eqnarray}
f(P) &=& \phi(P)Q\\
b_i &=& \lambda(P)
\end{eqnarray} where the function $\lambda$ has the following definition: \begin{displaymath}\lambda(P) = \left\{ \begin{array}{cc}
                                            1 & \textrm{if } \phi(P) \geq \frac{p+1}{2} \\
                                            0 & \textrm{otherwise}
                                          \end{array}
\right.\end{displaymath}

The domain of the Kaliski generator is $\mathcal{D} = E(\mathbb{F}_p)$ and the seed $P_1$ is a random point on the curve.

Suppose that a cryptosystem uses the Kaliski generator to produce pseudorandom quantities. This generator was initialized with the parameters $p= 5$ and $c = 1$. Suppose also that an adversary of this cryptosystem wants to attack a Kaliski generator.

In this scenario, suppose that the adversary ($i$) discovered that the following sequence of bits $\textbf{b} = 10$ was outputted by the generator; and, ($ii$) possess a quantum computer able to execute the generalized quantum permanent compromise attack to the Blum-Micali construction.

In the next section, details about the Kaliski generator under attack will be presented to the reader in order to clarify the comprehension about the steps of the attack. After that, the attack setup will be described, reporting all the gates and number of iterations required by the attack. To conclude the attack, the steps of the quantum algorithm will be detailed.

\section{Details of  Initialization of the Kaliski Generator Under Attack}
In the example of the Kaliski generator used in this file, the initialization adopted the parameters $p = 5$ and $c = 1$, resulting in the following equation of the curve:

\begin{equation}
y^2 = x^3 + 1 \bmod 5
\end{equation}

The set of points that satisfy this equation is $\left\{ (4,0), (0,1), (0,4), (2,2), (2,3) \right\}$. This set together with a point at infinity, denoted by $\mathcal{O}$, characterizes the cyclic group of order $p + 1$, i.e., the domain of the permutation.

The generator of this group is $Q = (2,2)$ and is important to remark that:

\begin{eqnarray}
Q &=& (2,2)\\
2Q &=& Q + Q = (0,4)\\
3Q &=& 2Q + Q = (4,0)\\
4Q &=& 3Q + Q = (0,1)\\
5Q &=& 4Q + Q = (2,3)\\
6Q &=& 5Q + Q = \mathcal{O}
\end{eqnarray} It is important to notice that $kQ$, where $k$ is an integer, does not represent the ordinary multiplication operation. It represents the addition of a point to itself in the context of an elliptic curve. More details about this operation should be seen in the book of Paar and Pelzl (Section $9.1.2$ -- Group Operations on Elliptic Curves) [7]  and also in the book of Stallings  (Section $6.5$ -- Elliptic Curves Over Finite Fields) [9].

The generator of the example has the form:

\begin{eqnarray}
P_{i} &=& \phi(P_{i-1})Q\\
b(P_i) &=& \lambda(P)
\end{eqnarray} where the function $\phi$ has the following definition:
\begin{displaymath}
\phi(P) = \left\{ \begin{array}{cc}
                    y & \textrm{if } P = (x,y)\\
                    p & \textrm{if } P = \mathcal{O}
                  \end{array}
\right.
\end{displaymath} The function $\lambda$ has the following definition: \begin{displaymath}\lambda(P) = \left\{ \begin{array}{cc}
                                            1 & \textrm{if } \phi(P) \geq 3 \\
                                            0 & \textrm{otherwise}
                                          \end{array}
\right.\end{displaymath}

\begin{figure}[h]
  \includegraphics[scale=0.23]{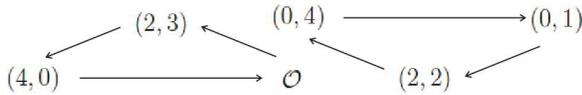}\\
  \caption{Functional graph for the one-way permutation of the Kaliski generator used in the example.}\label{fig:permutation}
\end{figure}

For this example, the resulting permutation can be represented as the functional graph illustrated in the Figure \ref{fig:permutation}.

\subsection{Attack Setup}
The attack setup comprehend all the steps necessary to prepare the quantum algorithm to run. Firstly is is necessary to determine how many qubits are necessary as input.

The number of qubits to represent the domain in a quantum computer is $\left\lceil \log \mathcal{D} \right\rceil = \left\lceil \log 6 \right\rceil = 3$. Since $2$ bits where discovered by the adversary, $2$ qubits will be necessary in the third register. In this way, the summarization of necessary qubits is: $3$ qubits to first register, $2$ qubits to the second register, and $1$ qubit as ancillary to the amplitude amplification procedure.

Since the points cannot be directly represented in a quantum computer, the following representation will be used:

\begin{eqnarray}
(4,0) &\equiv& \ket{1}\\
(0,1) &\equiv& \ket{2}\\
(0,4) &\equiv& \ket{3}\\
(2,2) &\equiv& \ket{4}\\
(2,3) &\equiv& \ket{5}\\
\mathcal{O} &\equiv& \ket{6}
\end{eqnarray}

The next step is to to prepare the quantum gates that will be used in the attack. The $\rho$ gate, responsible to implement the permutation, performs the following transformations:

\begin{eqnarray}
\ket{0} &\rightarrow& \ket{0}\\
\ket{1} &\rightarrow& \ket{6}\\
\ket{2} &\rightarrow& \ket{4}\\
\ket{3} &\rightarrow& \ket{2}\\
\ket{4} &\rightarrow& \ket{3}\\
\ket{5} &\rightarrow& \ket{1}\\
\ket{6} &\rightarrow& \ket{5}\\
\ket{7} &\rightarrow& \ket{7}
\end{eqnarray}  It should be noticed that the gate $\rho$ is unitary, since $\rho \cdot \rho^{\dagger} = \mathbb{I}$, where $\mathbb{I}$ denotes the identity matrix.

The gate $\lambda_0$ performs the following transformations:

\begin{eqnarray}
\ket{0}\ket{c} &\rightarrow& \ket{0}\ket{c}\\
\ket{1}\ket{c} &\rightarrow& \ket{1}\ket{\overline{c}}\\
\ket{2}\ket{c} &\rightarrow& \ket{2}\ket{\overline{c}}\\
\ket{3}\ket{c} &\rightarrow& \ket{3}\ket{c}\\
\ket{4}\ket{c} &\rightarrow& \ket{4}\ket{\overline{c}}\\
\ket{5}\ket{c} &\rightarrow& \ket{5}\ket{c}\\
\ket{6}\ket{c} &\rightarrow& \ket{6}\ket{c}\\
\ket{7}\ket{c} &\rightarrow& \ket{7}\ket{c}\\
\end{eqnarray}

%

In the case of the Kaliski generator, the matrix representation of the gates is shown in the Appendix \ref{apend:gates}. The reader can verify that they are unitary by performing a multiplication of each gate to it transpose conjugated.

The number of iterations required by the Grover's algorithm is given by:

\begin{equation}
k = \left\lfloor \frac{\pi}{4}\sqrt{\frac{8}{1}}\right\rceil = 2
\end{equation}

\begin{figure}[ht!]
\centering
\leavevmode
\scriptsize{
\Qcircuit @C=0.6em @R=1.0em {
&&&&&&&&&&&&&&&&&&&&&&&&&&&&&&&&&&&&&&\\
& & & & \lstick{\ket{0}}   & & \qw & \multigate{2}{H} & \qw & \multigate{3}{\lambda_0}   & \qw   & \qw  & \multigate{2}{\rho}   & \qw  &\multigate{4}{\lambda_0}  & \qw  & \multigate{2}{\rho} &\qw & \multigate{5}{G}    & \qw & \multigate{5.6]}{G} & \qw & \meter & \cw\\
& & & & \lstick{\ket{0}}   & & \qw & \ghost{H}        & \qw & \ghost{\lambda_0}          & \qw   & \qw  & \ghost{\rho}          & \qw  &\ghost{\lambda_0}         & \qw  & \ghost{\rho}        &\qw & \ghost{G}           & \qw & \ghost{G}        & \qw & \meter & \cw \\
& & & & \lstick{\ket{0}}   & & \qw & \ghost{H}        & \qw & \ghost{\lambda_0}          & \qw   & \qw  & \ghost{\rho}          & \qw  &\ghost{\lambda_0}         & \qw  & \ghost{\rho}        &\qw & \ghost{G}           & \qw & \ghost{G}        & \qw & \meter & \cw \\
& & & & \lstick{\ket{0}}   & & \qw & \qw              & \qw & \ghost{\lambda_0}          & \qw   & \qw  & \qw                   & \qw  &\ghost{\lambda_0}         & \qw  & \qw                 &\qw & \ghost{G}           & \qw & \ghost{G}        & \qw & \qw    & \qw \\
& & & & \lstick{\ket{0}}   & & \qw & \qw              & \qw & \qw                        & \qw   & \qw  & \qw                   & \qw  &\ghost{\lambda_0}         & \qw  & \qw                 &\qw & \ghost{G}           & \qw & \ghost{G}        & \qw & \qw    & \qw \\
\left\{ \right. & & & & \lstick{\ket{1}}   & & \qw & \gate{H} & \qw &  \qw               & \qw   & \qw  & \qw                   & \qw  & \qw                      & \qw  & \qw                 &\qw & \ghost{G}           & \qw & \ghost{G}        & \qw & \qw    & \qw\\
& & & &                    & &\nwx[-7] &              & \nwx[-7]&  & \nwx[-7]   &      &  & \nwx[-7]  &   & \nwx[-7]  &  & \nwx[-7]  & & & & \nwx[-7]           & &        &  &    &  \\
& & & &                    & &\ket{\psi_0} &              & \ket{\psi_1}&  & \ket{\psi_2}   &      &  & \ket{\psi_3}  &   & \ket{\psi_4}  &  & \ket{\psi_5}  & & & & \ket{\psi_6}           & &        &  &    &  \\
}}
\caption{Quantum circuit that implements the attack against the Kaliski generator.} \label{circ:total}
\end{figure}
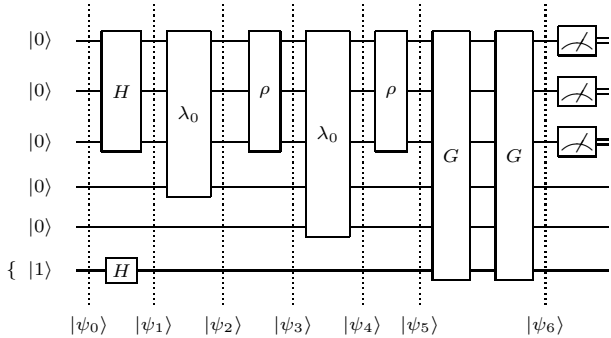

Arranging the gates as suggested by the algorithm, the resulting circuit is denoted in the Figure \ref{circ:total}.

\subsection{Attack Example}
The first step describes the initialization of the circuit according to each register as shown in the $\ket{\psi_0}$:

\begin{equation}
\ket{\psi_0} = \ket{000}\ket{00}\ket{1}
\end{equation}

It is applied to the first and third registers the Hadamard gate, responsible to put the input in an equally distributed superposition. The result of the application of such gate is shown in the $\ket{\psi_1}$:

{\small
\begin{eqnarray}
\ket{\psi_1} &=& H^{\otimes 3} \otimes I^{\otimes 2} \otimes H \ket{\psi_0}\\
&=& H^{\otimes 3}\ket{000} \ket{00} H\ket{1}\\
&=& \frac{1}{\sqrt{8}}\sum_{i=0}^{8} \ket{i}\ket{00}\ket{-}\\
&=& \frac{1}{\sqrt{8}} \left( \ket{0} + \ket{1} + \ldots \ket{7} \right) \ket{00} \ket{-}
\end{eqnarray}}

At this point, all the states have the same probability to be measured. The next step is to perform the first phase of the quantum permanent compromise algorithm, responsible for the identification of the representative. The $\lambda_0$ gate associate in the third register all the elements of the first one that would have produced the bit $0$ in the hard-core predicate. The result is shown in the $\ket{\psi_2}$ below:

{\footnotesize
\begin{eqnarray}
\ket{\psi_2} &=& \lambda_0 \ket{\psi_1}\\
&=& \frac{1}{\sqrt{8}}\left( \ket{1} + \ket{2} + \ket{4} \right)\ket{10}\ket{-} + \nonumber \\
&+& \frac{1}{\sqrt{8}}\left( \ket{0} + \ket{3} + \ket{5} + \ket{6} + \ket{7} \right)\ket{00}\ket{-}
\end{eqnarray}}

It is important to notice that up to this point the the candidates to the representative are: $\left\{ \ket{1}, \ket{2}, \ket{4}\right\}$. Since the algorithm reproduces the steps of the Kaliski generator, it is necessary to perform the permutation in all the elements of the domain. This operation is performed by the $\rho$ gate, as shown in the state $\ket{\psi_3}$.

{\footnotesize
\begin{eqnarray}
\ket{\psi_3} &=& \rho \ket{\psi_2}\\
&=& \frac{1}{\sqrt{8}}\left( \ket{6} + \ket{4} + \ket{3} \right)\ket{10}\ket{-} +\nonumber\\
&+& \frac{1}{\sqrt{8}}\left( \ket{0} + \ket{2} + \ket{1} + \ket{5} + \ket{7} \right)\ket{00}\ket{-}
\end{eqnarray}}

The next step is to apply again the gate $\lambda_0$, that will identify the elements that would have produced the second bit. The effect of this gate is reported in the $\ket{\psi_4}$.

{\footnotesize
\begin{eqnarray}
\ket{\psi_4} &=& \lambda_0 \ket{\psi_3}\\
&=& \frac{1}{\sqrt{8}}\ket{4}\ket{11}\ket{-} + \frac{1}{\sqrt{8}}\left(\ket{2} + \ket{1}\right) \ket{01}\ket{-} +\nonumber\\
&+& \frac{1}{\sqrt{8}} \left( \ket{6} + \ket{3} \right)\ket{10}\ket{-} +\nonumber \\
&+& \frac{1}{\sqrt{8}}\left( \ket{0} +  \ket{5} + \ket{7} \right)\ket{00}\ket{-}
\end{eqnarray}}

The next step is to perform the application of the gate $\rho$ one more time. It is necessary to identify the representant of the internal state $X(3)$.

{\footnotesize
\begin{eqnarray}
\ket{\psi_5} &=& \rho \ket{\psi_4}\\
&=& \frac{1}{\sqrt{8}}\ket{3}\ket{11}\ket{-} + \frac{1}{\sqrt{8}}\left( \ket{4} + \ket{6}  \right)\ket{01}\ket{-}\nonumber\\
&+& \frac{1}{\sqrt{8}} \left( \ket{5} + \ket{4} \right)\ket{10}\ket{-} + \nonumber\\
&+& \frac{1}{\sqrt{8}}\left( \ket{0} +  \ket{1} + \ket{7} \right)\ket{00}\ket{-}
\end{eqnarray}}

After that, it is important to notice that the representative of the internal state $X(3)$ is already identified: $\ket{3}$. However, a measurement in the second register at this point would return any number from $\ket{0}$ to $\ket{7}$ with the same probability. The next step of the algorithm comprehend the amplitude amplification of the element identified as solution. To proceed is necessary to consider the following representation of the state $\ket{\psi_5}$:

{\footnotesize
\begin{eqnarray}
\ket{\psi_5'} &=& \frac{1}{\sqrt{8}}\ket{3}\ket{11}\ket{-} + \frac{1}{\sqrt{8}}\sum_{j=0, j\ne 3}^{7}\ket{j}\ket{z \ne 11}\\
&=& \frac{1}{\sqrt{8}}\ket{\psi_{x_i}} + \sqrt{\frac{7}{8}}\ket{\psi_{\neg x_i}}
\end{eqnarray}} It should be noticed that there's a partition in two subspaces:$\ket{\psi_{x_i}} = \ket{3}\ket{11}\ket{-}$ and $\ket{\psi_{\neg x_i}} = \sum_{j=0, j\ne 3}^{7}\ket{j}\ket{z \ne 11}\ket{-}$.

Considering the geometric representation of this state, then:

\begin{equation}
\ket{\psi_5'}  = \sin{\theta} \ket{\psi_{x_i}} + \cos(\theta) \ket{\psi_{\neg x_i}}
\end{equation} where $\sin^2\theta = \frac{1}{8}$ and $\theta \in \left(0,\frac{\pi}{2}\right)$, therefore $\theta =  0.361$ radians.

The next step of the algorithm is to perform $k = 2$ Grover's iterations in the state $\ket{\psi_5'}$, resulting:

{\footnotesize
\begin{eqnarray}
\ket{\psi_6} &=& G^{\otimes 2} \ket{\psi_5'}\\
&=&\ sin[(2\cdot k + 1)\theta] \ket{\psi_{x_i}} + \nonumber \\
&+& \cos[(2\cdot k + 1)\theta] \ket{\psi_{\neg x_i}}\\
&=& \sin[5 \cdot 0.361] \ket{\psi_{good}} + \nonumber \\
&+& \cos[5 \cdot 0.361] \ket{\psi_{bad}}\\
&=& \sin(1.805)\ket{\psi_{good}} + \cos(1.805) \ket{\psi_{bad}}
\end{eqnarray}} At this point, a measurement in the second register would return the state $\ket{3}$ with probability of $\left| \sin(1.805) \right|^2 = 0.946$. With this information the intruder will be able to retrieve all the set $X(i)$ of internal states from the generator under attack, endangering its unpredictability.

This concludes an example of the generalized quantum permanent compromise attack against the security of the Kaliski generator.

\section{Final Remarks}
The examples illustrated in this file show how to endanger the security of the generators BBS and Kaliski from the Blum-Micali Construction. This endangering is made by a quantum permanent compromise attack and the consequence is that an adversary is capable to reproduce all the previous and future outputs of the generator.

The quantum attack is based on Amplitude Amplification, a generalization of Grover's quantum search. This attack provides a quadratic speedup over the classical analogous algorithm. For more details about the quantum attack, the reader is reported to the papers of Guedes et al. [3, 4].

\section*{Acknowledgements}
The authors gratefully acknowledge the financial support rendered by the Brazilian National Council for the Improvement of Higher Education (CAPES).

\appendix

\section{Matrix Representation of the Gates} \label{apend:gates}
\begin{equation*}
\rho = \left[ \begin{array}{cccccccc}
         1 & 0 & 0 & 0 & 0 & 0 & 0 & 0 \\
         0 & 0 & 0 & 0 & 0 & 0 & 1 & 0 \\
         0 & 0 & 0 & 0 & 1 & 0 & 0 & 0 \\
         0 & 0 & 1 & 0 & 0 & 0 & 0 & 0 \\
         0 & 0 & 0 & 1 & 0 & 0 & 0 & 0 \\
         0 & 1 & 0 & 0 & 0 & 0 & 0 & 0 \\
         0 & 0 & 0 & 0 & 0 & 1 & 0 & 0 \\
         0 & 0 & 0 & 0 & 0 & 0 & 0 & 1
       \end{array} \right]
\end{equation*}

{\footnotesize
\begin{equation*}
\lambda_0 = \left[ \begin{array}{cccccccccccccccc}
1 & 0 & 0 & 0 & 0 & 0 & 0 & 0 & 0 & 0 & 0 & 0 & 0 & 0 & 0 & 0\\
0 & 1 & 0 & 0 & 0 & 0 & 0 & 0 & 0 & 0 & 0 & 0 & 0 & 0 & 0 & 0\\
0 & 0 & 0 & 1 & 0 & 0 & 0 & 0 & 0 & 0 & 0 & 0 & 0 & 0 & 0 & 0\\
0 & 0 & 1 & 0 & 0 & 0 & 0 & 0 & 0 & 0 & 0 & 0 & 0 & 0 & 0 & 0\\
0 & 0 & 0 & 0 & 0 & 1 & 0 & 0 & 0 & 0 & 0 & 0 & 0 & 0 & 0 & 0\\
0 & 0 & 0 & 0 & 1 & 0 & 0 & 0 & 0 & 0 & 0 & 0 & 0 & 0 & 0 & 0\\
0 & 0 & 0 & 0 & 0 & 0 & 1 & 0 & 0 & 0 & 0 & 0 & 0 & 0 & 0 & 0\\
0 & 0 & 0 & 0 & 0 & 0 & 0 & 1 & 0 & 0 & 0 & 0 & 0 & 0 & 0 & 0\\
0 & 0 & 0 & 0 & 0 & 0 & 0 & 0 & 0 & 1 & 0 & 0 & 0 & 0 & 0 & 0\\
0 & 0 & 0 & 0 & 0 & 0 & 0 & 0 & 1 & 0 & 0 & 0 & 0 & 0 & 0 & 0\\
0 & 0 & 0 & 0 & 0 & 0 & 0 & 0 & 0 & 0 & 1 & 0 & 0 & 0 & 0 & 0\\
0 & 0 & 0 & 0 & 0 & 0 & 0 & 0 & 0 & 0 & 0 & 1 & 0 & 0 & 0 & 0\\
0 & 0 & 0 & 0 & 0 & 0 & 0 & 0 & 0 & 0 & 0 & 0 & 1 & 0 & 0 & 0\\
0 & 0 & 0 & 0 & 0 & 0 & 0 & 0 & 0 & 0 & 0 & 0 & 0 & 1 & 0 & 0\\
0 & 0 & 0 & 0 & 0 & 0 & 0 & 0 & 0 & 0 & 0 & 0 & 0 & 0 & 1 & 0\\
0 & 0 & 0 & 0 & 0 & 0 & 0 & 0 & 0 & 0 & 0 & 0 & 0 & 0 & 0 & 1\\
       \end{array} \right]
\end{equation*}}


\end{document}